\documentstyle[aps,twocolumn,graphicx,epsfig]{revtex}

\begin{document}


\draft

\title{Collective oscillations of two colliding Bose-Einstein
  condensates} \author{P.~Maddaloni\cite{pm}, M.~Modugno, C.~Fort,
  F.~Minardi, and M.~Inguscio} \address{INFM - European Laboratory for
  Non linear Spectroscopy (LENS) and Dipartimento di Fisica,
  Universit\`a di
  Firenze, \\
  Largo E. Fermi 2, I-50125 Firenze, Italy.}

\date{\today}

\maketitle

\begin{abstract}
  
  Two $^{87}$Rb condensates ($F=2, m_f$=2 and $m_f$=1) are produced in
  highly displaced harmonic traps and the collective dynamical
  behaviour is investigated. The mutual interaction between the two
  condensates is evidenced in the center-of-mass oscillations as a
  frequency shift of 6.4(3)\%. Calculations based on a mean-field
  theory well describe the observed effects of periodical collisions
  both on the center-of-mass motion and on the shape oscillations.

\end{abstract}

Since the first realization of Bose-Einstein condensation with dilute
trapped gases \cite{1stBEC}, systems of condensates in different
internal states have deserved attention as mixtures of quantum
fluids. In this context,
the important issue of the interaction between two distinct
condensates was early addressed at JILA \cite{jila1} with the
production of two $^{87}$Rb condensates in the hyperfine levels
$|F=2,m_f=2\rangle$ and $|1,-1\rangle$ in a Ioffe-type trap.
Subsequent experiments of the JILA group have focused on the dynamics
of two condensates in the states $|2,1\rangle$ and $|1,-1\rangle$
having nearly the same magnetic moment, confined by a time-orbiting
potential (TOP) trap. In these experiments \cite{jila2,jila3,jila4}
the authors have investigated the effects of the mutual interaction in
a situation of almost complete spatial overlap of the two condensates.
The resulting dynamics reveals a complex structure, and it is
characterized by a strong damping of the relative motion of the two
condensates \cite{jila2}.  More recently another group \cite{otago}
has experimentally investigated a mixture of $^{87}$Rb condensates in
different $m_f$ states in a TOP trap, but no effects of the mutual
interaction have been observed.

These experiments have raised several interesting questions from the
theoretical point of view, such as the origin of the above mentioned damping
\cite{sinatra}, the phase coherence properties of the two condensates
\cite{sinatra2} and the nature of Rabi oscillations in presence of an
external coupling \cite{williams}, thus providing a challenge to
explore new dynamical regimes in different experimental
configurations.

In this work we demonstrate an experimental method for a sensitive and
precise investigation of the interaction between two condensates.
These are made to collide after periods of spatially separated
evolution and we get quantitative information from the resulting
collective dynamics.  Following the scheme originally introduced by
W.~Ketterle and co-workers \cite{alK} to create an atom laser
out-coupler, we use a radio-frequency (rf) pulse to produce two
$^{87}$Rb condensates in the states $|2,2\rangle \equiv |2\rangle$ and
$|2,1\rangle \equiv |1\rangle$. Due to the different magnetic moments
and the effect of gravity, they are trapped in two potentials whose
minima are displaced along the vertical $y$ axis by a distance much
larger that the initial size of each condensate. As a consequence the
$|1\rangle$ condensate, initially created in the equilibrium position
of $|2\rangle$, undergoes large center-of-mass oscillations, in a
regime very different from that explored in \cite{jila2} and analyzed
in \cite{sinatra}, where the two condensates sit in overlapping traps.
The fact that the two condensates periodically collide opens the
possibility of detecting even small interactions through changes in
frequency and amplitude of the oscillations. Indeed, the periodic
collisions of the $|1 \rangle$ condensate with the $|2\rangle$,
initially remaining almost at rest, strongly affect the collective
excitations of both the condensates: {\it (i)} the center-of-mass
oscillation frequency of the $|1\rangle$ condensate is shifted
upwards; {\it (ii)} the shape oscillations of $|2\rangle$ condensate,
triggered by the sudden transfer to the $|1\rangle$ state, are
significantly enhanced.

The complex dynamics is quantitatively analyzed and found in agreement
with the theoretical predictions derived by the numerical solution of
two coupled Gross-Pitaevskii (GP) equations at zero temperature.

We prepare a condensate of typically $1.5 \times 10^5$ $~^{87}$Rb
atoms in the $F=2,m_f=2$ hyperfine level ($|2\rangle$), confined in a 4-coils
Ioffe-Pritchard trap elongated along the $z$ symmetry axis \cite{epj}.
The axial and radial frequencies for the $|2\rangle$ state,
measured after inducing center-of-mass oscillations of the condensate,
are $\omega_{z 2}= 2 \pi \times 12.6(2)$~Hz and $\omega_{\perp 2}= 2
\pi \times 164.5(5)$~Hz respectively, with a magnetic field minimum of
$1.75$~Gauss. By applying a rf pulse, the initial $|2\rangle$
condensate is put into a coherent superposition of different Zeeman
$|m_f\rangle$ sublevels of the $F=2$ state, which then move apart: $|2
\rangle$ and $|1 \rangle$ are low-field seeking states and stay
trapped, $|0 \rangle$ is untrapped and falls freely under gravity,
while $|-1 \rangle$ and $|-2 \rangle$ are high-field seeking states
repelled from the trap.  All the condensates in different Zeeman
states are simultaneously imaged by absorption with a 150~$\mu$s pulse
of light resonant on the $F=2 \rightarrow F'=3$ transition, shone
30~ms after the switching-off of the trap. By fixing the duration and
varying the amplitude of the rf field, we control the relative
population in different Zeeman sublevels \cite{Erice}.  A 10 cycles rf
pulse at 1.24~MHz with an amplitude B$_{rf}$=10~mG quickly transfers
$\sim 13$\% of the atoms to the $|1\rangle$ state without populating
the $|0\rangle$, $|-1 \rangle$ and $|-2 \rangle$ states. The
$|1\rangle$ condensate experiences a trapping potential with lower
axial and radial frequencies ($\omega_1=\omega_2 / \sqrt{2}$) and
displaced along the vertical $y$ axis by $y_0 = g/ \omega_{\perp 2} ^2
\simeq 9.2$~$\mu$m.  After the rf-pulse, the $|1\rangle$ condensate
moves apart from $|2\rangle$, and begins to oscillate around its
equilibrium position. Due to the mutual repulsive interaction, the
latter starts oscillating too, though with a much smaller amplitude.
The resulting periodic superimposition modifies the effective
potential, which is the sum of the external potential (magnetic and
gravitational) and the mean-field one.  We have studied the dynamics
of the $|1\rangle$ condensate in presence (``interacting'' case) and
in absence (``non-interacting'' case) of $|2\rangle$, by varying the
permanence time in the trap.  We restrict to permanence times so
short, i.e. less than 40~ms, that we can neglect both atom losses due
to the condensate finite lifetime, 0.7(1)~s, and the heating
($dT/dt=0.11(2)~\mu$K/s).  For the non-interacting case, we have used
a stronger rf-pulse (B$_{rf}$=20~mG) that completely empties the
$|2\rangle$ state coupling all the atoms in $|1\rangle$ and in the
other untrapped Zeeman states, which rapidly leave the trap.

\begin{figure}[ht]
\epsfig{file=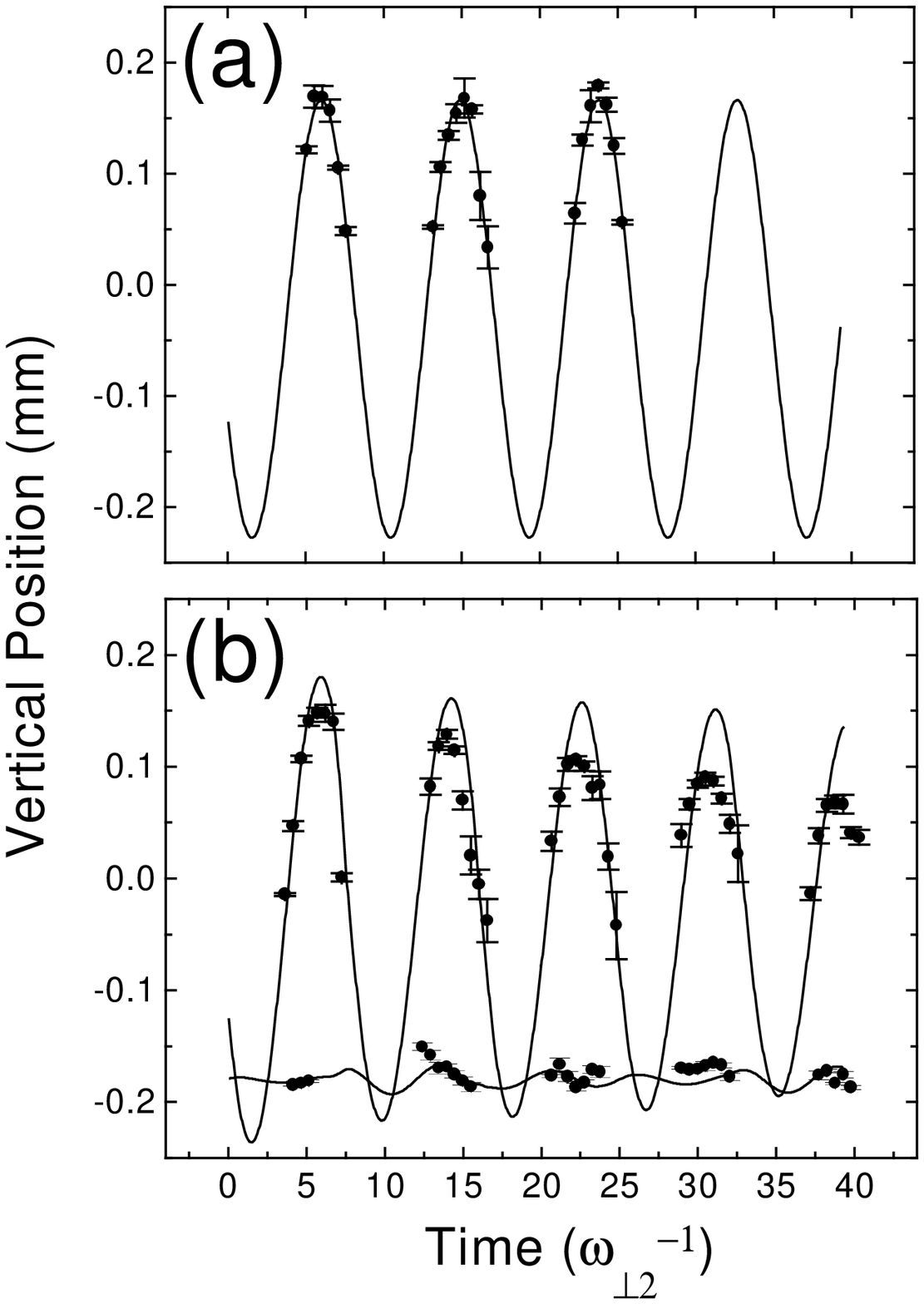,width=8cm}
\caption{Center-of-mass oscillations as a function of the trapped
  evolution time (rf-pulse at $t=0$) after 30~ms of ballistic
  expansion for (a) the non-interacting $|1\rangle$ condensate and (b)
  the two interacting condensates. The experimental points are
  compared with the relative theoretical predictions (discussed in the
  text).  Each data point is the average of typically five
  measurements.}
\label{fig:cyexp}
\end{figure}

We start considering the case of the $|1\rangle$ condensate in absence
of the $|2\rangle$ condensate (non-interacting case). In
Fig.~\ref{fig:cyexp}a we plot the center-of-mass oscillations as a
function of the trapped evolution time in units of
$\omega_{\perp2}^{-1}\simeq 0.967$~ms, after 30~ms of ballistic
expansion. The center-of-mass undergoes sinusoidal oscillation with a
measured frequency of $\omega_{\perp1}=2 \pi \times 116.4(3)$~Hz. This
value is, as expected, a factor $\sqrt{2}$ smaller than
$\omega_{\perp2}$.  Furthermore the experimental data give no evidence
of damping within the maximum monitored trapping time (26~ms).
Fig.~\ref{fig:cyexp}b shows the center-of-mass evolution of the two
condensates in the interacting case. The center-of-mass of the
$|1\rangle$ condensate exhibits a substantially different behaviour,
which allows a clear quantitative analysis. The oscillation frequency
is up-shifted to $\omega_{\perp 1}=2\pi \times 123.9(3)$~Hz, i.e. a
6.4(3)\% change with respect to the non-interacting case.  This can be
understood considering that the mean-field repulsion of the second
condensate produces an anharmonic correction to the effective
potential experienced by the $|1\rangle$ condensate. Furthermore, the
oscillations appear now damped with an exponential decay time of about
60~ms. Indeed, each time the two condensates superimpose (about every
8~ms), there is an energy transfer from the
$|1\rangle$ condensate toward the $|2\rangle$ condensate. As a
consequence we expect, and we do observe, effects on both the
center-of-mass motion and the collective shape-oscillations of the
$|2\rangle$ condensate.

A quantitative description of these experimental features requires the
solution of two coupled Gross-Pitaevskii (GP) equations.  Neglecting
the interaction with the thermal cloud, the two condensates evolve
according to
\begin{equation}
i\hbar{\partial\Psi_i\over\partial t} = \left[-
{\hbar^2\nabla^2\over2m} + V_i
+\sum_{j=1,2} {4\pi\hbar^2 a_{ij}\over m}|\Psi_j|^2 \right]\Psi_i
\label{eq:gp0}
\end{equation}
$i=1,2$, where $V_i$ are the trapping potentials:
\begin{eqnarray}
V_1(x,y,z) &=& {m\over2}\omega_{\perp1}^2 \left[(x^2 + y^2)
+ \lambda^2 z^2\right]\\
V_2(x,y,z) &=& {m\over2}\omega_{\perp2}^2 \left[\left(x^2 +
    (y-y_0)^2\right)
+ \lambda^2 z^2\right]
\end{eqnarray}
and the asymmetry parameter is 
$\lambda\equiv\omega_{z}/\omega_{\perp}\simeq~0.0766$ for both traps.
For the $^{87}$Rb scattering
lengths we use $a_{22}=a_{12}=98\,a_0$ and $a_{11}=94.8\,a_0$
\cite{nist}.  

Our experimental configuration allows to simplify these equations by
using the fact that we are in the Thomas-Fermi (TF) regime due to the
large number of atoms, and that the system is strongly elongated along
the $z$ axis, $\lambda\ll 1$ \cite{forthcoming}. For an elliptic
trap, the low-frequency excitations with $m=0$ are linear
superpositions of the monopole ($n=1$, $l=0$, $m=0$) and quadrupole
($n=0$, $l=2$, $m=0$) modes \cite{stringari,bec_review}. The
dispersion laws for the two modes, at leading order in $\lambda$, are
given by
\begin{equation}
\omega_{high}\simeq2 ~\omega_{\perp};
\qquad\omega_{low}\simeq \sqrt{5\over2} ~\lambda~\omega_{\perp}.
\end{equation}
In this limit the two frequencies are quite different, and the axial
and radial collective excitations are almost decoupled. The radial
width is characterized by small-amplitude oscillations with frequency
$\omega_{low}$ modulated by a large-amplitude oscillation with
frequency $\omega_{high}$, and vice versa for the axial width
\cite{ketterle}. Therefore, since the interactions mostly affect the
radial motion, we assume the axial dynamics to be still
characterized by the low frequency oscillations of the TF regime.
Then, we study the trapped dynamics in the $x,y$ plane by solving the
GP equation (\ref{eq:gp0}), by approximating our system as a uniform
cylinder \cite{numerical}.

We start from an initial configuration corresponding to the stationary
ground-state of Eq.\ (\ref{eq:gp0}), with all the $N$ trapped atoms in
the $|2\rangle$ condensate.  Afterwards, at $t=0$, $N_1$ atoms are
instantaneously transferred from the $|2\rangle$ to the $|1\rangle$
state, the former remaining with $N_2=N-N_1$ atoms. Here we consider
$N_1=0.13 N$ for the interacting case, and $N_1=N$ for the
non-interacting one.

The theoretical curves in Fig.~\ref{fig:cyexp} show an up-shift
of the center-of-mass oscillation frequency for the $|1\rangle$
condensate occurring in the interacting case, as experimentally
observed. This shift is of
5.4\%, in qualitative agreement with that measured. Furthermore, in
presence of interactions the
model correctly predicts a damping, which is not due to any
dissipative process (the total energy is conserved), but to a transfer
of energy from the center-of-mass oscillations of the $|1\rangle$
condensate to the other degrees of freedom of the system
\cite{SmeDal}. Still, this damping time is nearly a factor 2
longer than that experimentally observed. 

To understand the origin of the discrepancies between theory
and experiment, it's worth discussing the main approximations of our
model. First, since the model is basically 2-dimensional, the energy
transfer in the axial direction is overlooked.  Secondly, we
completely disregard the interaction between the two condensates
during the expansion \cite{tf}. Indeed, in our experiment, during the
switching-off of the trap the two clouds acquire different velocities
and cross each other in the fall, but we expect that they are so
dilute that we can neglect their mutual mean-field repulsion. Finally,
our model doesn't take into account the elastic scattering, occurring
when the relative velocity of the two condensates exceeds the sound
velocity \cite{nist:el_sc}. This effect, whose description lies beyond
the mean-field approximation, represents an important channel of both
atoms and energy losses which plays a significant role, for example, in the
four-wave mixing experiments with Bose condensates \cite{4wm,nist:pr}.

\begin{figure}
\epsfig{file=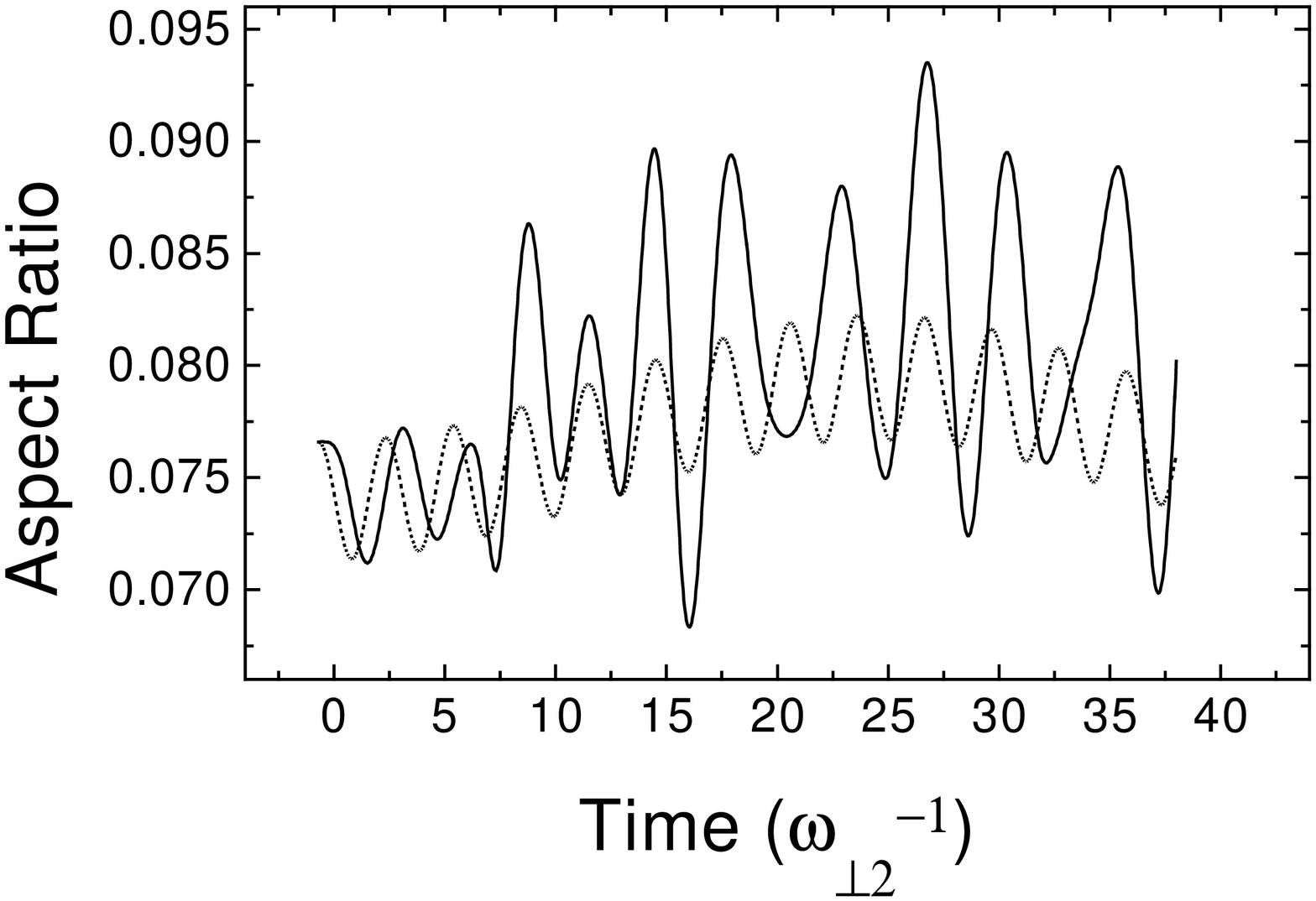,width=8cm}
\caption{Evolution of aspect ratio for the $|2\rangle$
  condensate in the trap obtained solving the GP equations for the interacting (solid line)
  and non-interacting case (dashed line). In the interacting case the
  evolution of the aspect ratio is not a simple superposition of the
  two oscillations with frequency $\omega_{low}$ and $\omega_{high}$,
  as in the non-interacting case.  We note also that the interaction
  between the two condensates produces an initial delay for the onset
  of collective excitations, since the two condensates are created
  with the same density profile, and the scattering lengths $a_{22}$
  and $a_{12}$ have the same value.}
\label{fig:artheo}
\end{figure}

\begin{figure}[ht]
\epsfig{file=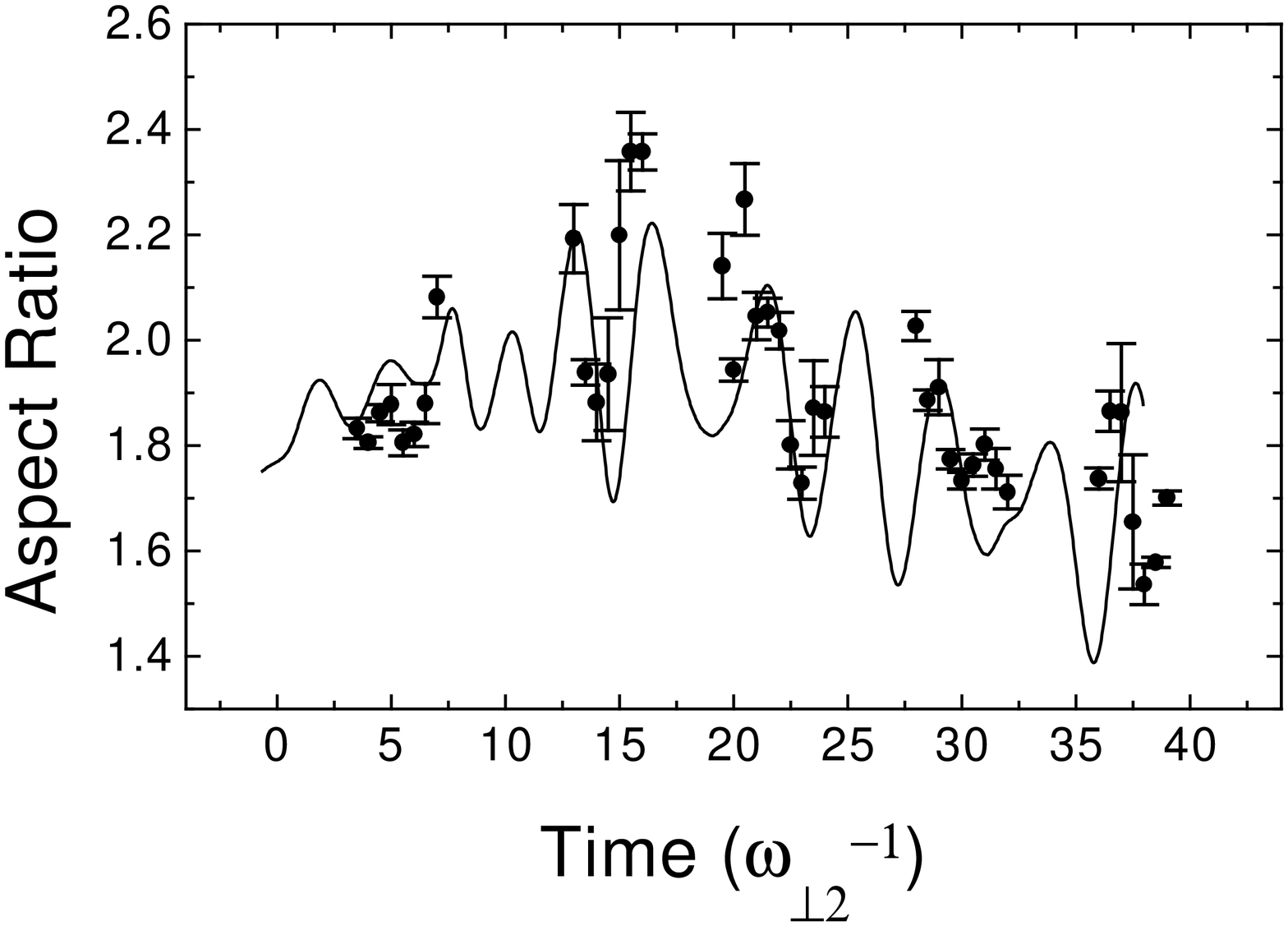,width=8cm}
\caption{Aspect ratio for the $|2\rangle$ condensate, 
  as a function of the trapped evolution time, after  30~ms of ballistic
  expansion. The experimental points are compared with the
  theoretical predictions for the case of two interacting condensates.
  }
\label{fig:arexp}
\end{figure}

We consider now the $|2\rangle$ condensate. In the interacting case,
both small center-off-mass oscillations and significant features for
the aspect ratio oscillations emerge from our model. The latter are
initially induced by the sudden change in the internal energy,
consequent on the transfer of $N_1$ atoms from the $|2\rangle$ to the
$|1\rangle$ level \cite{alK,jila5}. In Fig.~\ref{fig:artheo} we
compare the theoretical evolution of the $|2\rangle$ condensate aspect
ratio, i.e. the radial to axial width ratio, in absence and in
presence of collisions. In the former case, a faster (radial)
oscillation superimposes to a slow (axial) one as a consequence of the
decoupling between the two oscillation modes. The frequencies
$\omega_{high}$ and $\omega_{low}$ were separately measured by means
of resonant modulation of the trapping magnetic field \cite{epj}.  The
non-interacting behaviour predicted in Fig.~\ref{fig:artheo} should of
course yield when there is only one trapped state. For instance this
is the case of an atom laser out-coupled in a single-step transition,
as for sodium \cite{alK}, or for rubidium in $F=1$ state \cite{bloch}.
In the interacting case, we see from Fig.~\ref{fig:artheo} that a
significantly different behaviour is expected. Before the first
collision at $t \simeq 8$~ms, the oscillations are rather small since
they are determined only by the sudden change in the number of atoms.
Instead, at longer times, the changes in the aspect ratio become more
pronounced due to the energy transfer during collisions between the
two condensates.  Once the ballistic expansion is taken into account,
the simulation results are compared to the experimental data in
Fig.~\ref{fig:arexp}. The agreement is only of a qualitative
character. We attribute the discrepancy to the approximations we used
in our simplified analysis. Nevertheless, we believe that our model
provides a useful physical insight of the relevant aspects of the
problem.

In conclusion we have developed a powerful tool for the investigation
of the interactions between condensates and we have demonstrated how
they can quantitatively affect frequency, amplitude and shape of
oscillations. In particular, 
the frequency shift measurement gives
access to the number of atoms in the parent condensate $N_2$ or,
alternatively, to the $a_{21}$ scattering length. This
opportunity seems
particularly interesting for low $N_2$, as we have found that the
frequency shift scales roughly as
$\log(N_2a_{12})$ \cite{forthcoming}.

The agreement with the model, given its simplicity and lack of free
parameters, is generally satisfactory.  At least the discrepancy
observed for the damping of center-of-mass oscillations suggests that
the investigation of the relaxation beyond the mean-field theory
(namely, by including the elastic scattering) could be an interesting
subject for future studies.  The experimental perspectives of this
system of two periodically overlapping condensates lead to the
investigation of the interference between condensates spatially
separated during their evolution and, eventually, to studies of
Josephson effects in two weakly linked condensates \cite{smeJos}.

We acknowledge fruitful discussions with F.~Dalfovo.



\begin{references}
\bibitem{numerical} We solve the GP equations by using a Split-Step
  Fourier method \cite{jackson}, with a FFT algorithm \cite{recipes}.
\bibitem{tf} We numerically solve the differential equations for the
  scaling parameters in the Thomas-Fermi regime
  \cite{bec_review}.
\bibitem{nist:pr} Motivated by our work, preliminary 
  simulations, carried out by the NIST group in Gaithersbourg, show
  that including the elastic scattering greatly improves the
  comparison with the experimental damping, while the energy transfer in
  the axial direction is almost negligible: J. Burke, Y. Band and P.
  Julienne, private comunication.
\end{references}
\end{document}